\documentclass[fleqn,twoside]{article}
\usepackage[headings]{espcrc2}

\readRCS
$Id: espcrc2.tex,v 1.2 2004/02/24 11:22:11 spepping Exp $
\usepackage{graphicx}
\usepackage[figuresright]{rotating}

\title{Recent Work on Gravitational Waves From a Generic Standard Model-like Effective Higgs Potential}

\author{John Kehayias\address{Physics Department, University of California Santa Cruz, \\ 
        Santa Cruz, CA, US}
        \thanks{kehayias@physics.ucsc.edu}}
       
\runtitle{}
\runauthor{John Kehayias}

\begin{document}

\begin{abstract}
  I present recent work on gravitational waves (GWs) from a generic
  Standard Model-like effective potential for the electroweak phase
  transition.  We derive a semi-analytic expression for the
  approximate tunneling temperature, and analytic and approximate
  expressions for the two GW parameters $\alpha$ and $\beta$.  A quick
  summary of our analysis and general results, as well as a list of
  some specific models which easily fit into this framework, are
  presented.  The work presented here has been done in collaboration
  with Stefano Profumo \cite{paper}.  \vspace{1pc}
\end{abstract}

\maketitle

\section{Introduction}
As the temperature is lowered in the finite temperature quantum field
theory description of the electroweak Higgs sector (for a review, see
e.g.~\cite{tempreview}), it is possible to have a first order phase
transition through quantum mechanical tunneling.  A degenerate vacuum
state develops at $T_c$, and what began as the true vacuum of the
theory can become unstable at a lower temperature $T_{dest}$.  A
potential barrier separates this state from the true vacuum, and
tunneling to the lower energy state is probable at a temperature
$T_t$, where $T_{dest} \le T_t < T_c$.

Gravitational waves (GWs) can arise from a strongly first-order phase
transition through both turbulence and bubble nucleation.  Here, a
bubble is an area of the universe which has transitioned to the true,
electroweak symmetry breaking, vacuum.  Some fraction of the energy
released in these processes gives rise to a stochastic background of
gravitational waves.

It has been known for some time that the electroweak phase transition
in the minimal version of the Standard Model (SM) is not strongly
first order, given the experimental bounds on the Higgs mass.
However, many models of physics beyond the SM, including
supersymmetry, can enhance the phase transition and produce a GW
spectrum which might be experimentally observed in the near future.
While many models have been studied extensively, there has not been
much work done on a general, model-independent analysis.

In this note I will very briefly summarize work soon to be submitted
on studying generic effective potentials for the electroweak phase
transition.  The potential is very similar in form to the SM Higgs
potential, and the general results are applicable to several models
beyond the SM.

\section{Analysis of a Generic Effective Higgs Potential}
We consider a potential for the Higgs which mirrors that of the
(finite temperature, one loop, high temperature expansion) SM case of
the following form:
\begin{equation}\label{eq:genpot}
  V_{eff}(\phi,T) = \frac{\lambda(T)}{4}\phi^4 - (ET - e)\phi^3 + D(T^2 - T_0^2)\phi^2.
\end{equation}
In the SM $e = 0$.  

A semi-analytic expression for the three dimensional Euclidean action,
which is the important quantity for finite temperature tunneling, was
found in \cite{se3approx}.  The tunneling temperature, $T_t$, is
defined as the temperature when the probability to nucleate a bubble
in a horizon volume is $\mathcal{O}(1)$, a condition that is well
approximated by
\begin{equation}
  S_{E3}/T_t \sim 140,
\end{equation}
where we have assumed the temperature scale is
$\mathcal{O}(100\textrm{GeV})$ (see, e.g.~\cite{ewgravwave}).  Using
the results of \cite{se3approx}, we can thus derive an approximation
for $T_t$.

At $T_c$ for our potential the expression for $S_{E3}$ from
\cite{se3approx} has a singularity.  It also decreases very rapidly as
the temperature is lowered, to $0$ at $T_0$.  This implies that $T_t$
will be very close to $T_c$, and so we expand in powers of $\epsilon$:
$T \rightarrow T_c - \epsilon$.  The final lowest order expression for
$S_{E3}/T$ has all the parameters of the potential and is proportional
to $1/\epsilon^2$.  The singularity as $\epsilon \rightarrow 0$
remains, and $\epsilon$ is solved for by setting $S_{E3}/T = 140$.
Our approximation for the tunneling temperature is then
\begin{equation}
  T_t \approx T_c - \epsilon,
\end{equation}
with $\epsilon \ll 1$.

From our potential it is possible to calculate the exact GW
parameters, $\alpha$ and $\beta$.  $\alpha$ characterizes the energy
change of the vacuum transition, while $\beta$ characterizes the
bubble nucleation rate per unit volume.  These are evaluated at $T_t$,
which we now have an approximation for, but they are rather lengthy
expressions.  However, from our approximation for $S_{E3}/T$, a simple
expression for $\beta$ is obtained, which, to lowest order, is
proportional to $(T_c - \epsilon)/\epsilon^3$.

\section{The Parameter Space and Models}
We enforce that the potential of eq.~(\ref{eq:genpot}) describes
electroweak symmetry breaking with a Higgs boson.  This constrains the
vev of $\phi$ to be the usual $v \approx 246\textrm{ GeV}$, which must
be a stable minimum, and furthermore that the mass of the Higgs is
above the current experimental bound of $114\textrm{ GeV}$.  The signs
of the parameters (except for $e$) are fixed through this and the
potential considered in \cite{se3approx} (for general stability,
etc.), and we then also have $T_0^2 = v(3e + \lambda v)/2D$, which is
similar to the SM form.  There are constraints on $e$ based on its
sign, and $\lambda$ is set based on $e, v,$ and the Higgs mass $m_h$.
We also want the theory to be perturbative, so $\lambda < 1$, giving us
a mass range of $115\textrm{ GeV} \le m_h < 348\textrm{ GeV}$ (set by
the SM case of $0.11 \le \lambda < 1$).

Besides varying $\lambda$ we also vary one other parameter at a time.
For parameters besides $e$, which has constraints on its range, we
vary each up to two orders of magnitude larger and smaller than the SM
value.  In plotting the $\alpha-\beta$ plane we describe how the
various terms in the potential affect the GW spectrum parameters.
This covers as much as $12$ orders of magnitude for both $\alpha$ and
$\beta$.  The most remarkable enhancement for $\alpha$, which would
greatly contribute to observing a signal with future experiments,
comes at $e < 0$.

Several models can provide changes to the parameters in the potential
from their SM values.  There has been much work
(e.g.~\cite{delaunay,mssm}) on adding non-renormalizable terms to the
SM or MSSM which can enhance $E$, adding to the strength of the phase
transition.  The addition of an $SU(2)_L$ triplet (see,
e.g.~\cite{triplet}) provides a contribution to $\lambda$.  Since we
find that the additional parameter $e$ can greatly enhance $\alpha$,
models which add this term to the effective potential are particularly
interesting.  One such model is the addition of a gauge singlet,
arising as a solution to the $\mu$ problem in supersymmetry, for
instance.  The phenomenology of this model has been studied
extensively (e.g.~\cite{singlet}), and it has been shown that it is
possible to find evidence of such an additional singlet at the LHC,
allowing us to draw an interesting connection between collider and GW
physics.

\end{document}